\documentclass[conference]{IEEEtran}
\usepackage{cite}
\usepackage{amsmath,amssymb,amsfonts}
\usepackage{algorithm}
\usepackage{algpseudocode}
\usepackage{graphicx}
\usepackage{textcomp}
\usepackage{xcolor}
\usepackage{breqn}
\usepackage{amsmath,amssymb}
\usepackage{url}
\usepackage{subcaption}
\usepackage{enumitem}
\usepackage{comment}
\usepackage{mathtools}
\usepackage{multirow}
\usepackage{balance} 

\def\BibTeX{{\rm B\kern-.05em{\sc i\kern-.025em b}\kern-.08em
    T\kern-.1667em\lower.7ex\hbox{E}\kern-.125emX}}
\begin{document}

\newtheorem{theorem}{Theorem}
\newtheorem{lemma}{Lemma}
\newtheorem{assumption}{Assumption}
\newtheorem{definition}{Definition}

\newcommand{\der}{\mathrm{d}}
\newcommand{\cS}{\mathcal{S}}
\newcommand{\cV}{\mathcal{V}}
\newcommand{\cG}{\mathcal{G}}
\newcommand{\cE}{\mathcal{E}}
\newcommand{\cM}{\mathcal{M}}
\newcommand{\cN}{\mathcal{N}}
\newcommand{\cT}{\mathcal{T}}
\newcommand{\bu}{\mathbf{u}}    
\newcommand{\bw}{\mathbf{w}}   
\newcommand{\bv}{\mathbf{v}}    
\newcommand{\bx}{\mathbf{x}}
\newcommand{\by}{\mathbf{y}}
\newcommand{\bq}{\mathbf{q}}
\newcommand{\br}{\mathbf{r}}
\newcommand{\bd}{\mathbf{d}}

\newcommand{\mR}{\mathbb{R}}    
\newcommand{\mZ}{\mathbb{Z}}    
\newcommand{\va}{\vec{a}}     
\newcommand{\vb}{\vec{b}}

\newcommand{\tf}{\ensuremath{F}}
\newcommand{\tq}[1]{\ensuremath{q_{#1}}}

\newcommand{\defeq}{\mathrel{\mathop:}=}
\newcommand{\lat}[2]{\ensuremath{{\omega}\left(#1, #2  \right)}}
\newcommand{\ener}[2]{\ensuremath{{E}\left(#1, #2 \right)}}
\newcommand{\bud}[1]{\mathbf{u}^{#1}}
\newcommand{\uhat}{\mathbf{\hat{u}}} 
\newcommand{\bzd}[1]{\mathbf{z}^{#1}} 
\newcommand{\proj}[1]{\text{Proj}_{\Omega_c}\left(#1\right)}
\newcommand{\projall}[1]{\text{Proj}_{\Omega}\left(#1\right)}
\newcommand{\offload}[2]{{\tau(#1,#2)}}
\newcommand{\window}{\Delta}

\title{A Continuum Approach for Collaborative Task Processing in UAV MEC Networks}

\author{\IEEEauthorblockN{Lorson Blair, Carlos A. Varela, Stacy Patterson}
\IEEEauthorblockA{\textit{Department of Computer Science} \\ 
\textit{Rensselaer Polytechnic Institute} \\
Troy, NY, USA \\
blairl@rpi.edu, cvarela@cs.rpi.edu, sep@cs.rpi.edu}
}
\maketitle
    
\begin{abstract}
Unmanned aerial vehicles (UAVs) are becoming a viable platform for sensing and estimation in a wide variety of applications including disaster response, search and rescue, and security 
monitoring. These sensing UAVs have limited battery and computational capabilities, and thus must offload their data so it can be processed to provide actionable intelligence. We consider a compute platform consisting of a limited number of highly-resourced UAVs that act as mobile edge computing (MEC) servers to process the workload on premises. We propose a novel 
distributed solution to the collaborative processing problem that adaptively positions the MEC UAVs in response to the changing workload that arises both from the sensing 
UAVs' mobility and the task generation. Our solution consists of two key building blocks: (1) an efficient workload estimation process by which the UAVs estimate the task 
field---a continuous approximation of the number of tasks to be processed at each location in the airspace, and (2) a distributed optimization method
 by which the UAVs partition the task field so as to maximize the system throughput. We evaluate our 
proposed solution using realistic models of surveillance UAV mobility and show that our method achieves up to  28\% improvement in throughput over a non-adaptive baseline approach.
\end{abstract}

\begin{IEEEkeywords}
Mobile edge computing, task offloading, Voronoi partitioning.  
\end{IEEEkeywords}

\section{Introduction}

In its 2021 climate change report, the United Nations' International Panel on Climate Change (IPCC) gave a grim forecast regarding the frequency, severity, and devastating consequences
of natural disasters---hurricanes, wild fires, floods, etc.---as a result of climate change~\cite{IPCC2021Chap11}. With more catastrophic disasters, there is an urgent need for more 
sophisticated prediction models and enhanced disaster management and search and rescue techniques. Due to their flexible mobility, relatively cheap cost, and ease of deployment, unmanned aerial 
vehicles (UAVs) are becoming increasingly popular for use in these applications~\cite{Hayat-survey,mozaffari-uavapps}.
UAVs can be quickly and 
efficiently deployed in disaster areas to assess the level of damage, monitor the progression of wildfires and floods, draw digital maps of the disaster area, and
identify rescue candidates~\cite{uav3dmodeling}.

These applications require the collection of large amounts of data, for example, video, infrared imaging, or air quality data, which must be processed to generate actionable intelligence.
UAVs have limited physical space, which is reserved for data collection hardware, as well as a limited battery capacity that must be dedicated to this data collection. Thus, the UAVs, or \emph{mobile sensing agents} (MSAs), must offload data processing tasks to external resources. While a traditional solution is to offload this computation to the cloud, the urgency of the situation and potentially compromised infrastructure require offloading to nearby compute resources for immediate processing. 
A promising approach is to utilize a compute platform consisting of additional vehicles equipped with larger batteries and dedicated Mobile Edge Computing (MEC) servers~\cite{li-et-al,vfc-matching,uav-wpt,zhao-multi-agent-drl,liao-hotspot,online-uav-mounted}. 
These \emph{mobile compute agents} (MCAs) can be UAVs with larger batteries and specialized computing hardware, or they can be ground vehicles with their own power sources 
and servers. 
The MCAs have the freedom and flexibility to move in close proximity to the MSAs and to change locations in response to demand that changes both in location and quantity.

The successful deployment of such a platform requires addressing several research challenges. First, both the locations and quantities of computing demand are unpredictable and dynamic; however, this information is crucial for determining the best  placement of the MCAs. Second, even if the demand is known, the problem of optimally adapting the locations of the MCAs is intractable on its face due to the complexity of wireless communication and the binary nature of associating tasks to MCAs~\cite{huang-opt,yang-load-bal}. 
Third, unlike offloading to the cloud, MCAs have limited processing capacity, and so the assignment of tasks must also take these capacities into account to ensure balanced task processing. Finally, in disaster scenarios, communication infrastructure may be severely limited due to damage or the remoteness of the location; thus, the MCAs must be able to autonomously coordinate to provide the task processing services, reserving external communication for important results.

Several early works proposed solutions for an MEC platform consisting of a single UAV~\cite{li-et-al,joint-offloading,zhou-compratemax,noma-6g,energy-aware,uav-wpt}; however, a single UAV may not be sufficient to meet the computing demands of a large number of MSAs. More recent works have considered the deployment of multi-UAV MEC systems and address the problem of optimizing the MEC UAV trajectories for performance measures like latency and energy.
Many of these works assume that the demand is stationary~\cite{wang-ofld-drl,huang-opt,zhao-multi-agent-drl,yang-load-bal}.
A few works support dynamic demand~\cite{yan-data,liao-hotspot,online-uav-mounted}. 
However, all of these works propose centralized solutions, making them unsuitable for our setting.

We propose a novel distributed solution to maximize processing throughput in a multi-UAV MEC system. 
Our solution adaptively positions a limited number of MCAs in response to the changing workload that arises both from the mobility and task generation of the MSAs.
We adopt a \emph{continuum} approach to model the distribution of demand over the surveillance region. This unique approach enables us to formulate a continuous optimization problem that admits a distributed solution.
Our solution consists of two key building blocks:
(1)~an efficient demand estimation algorithm in which the MCAs collaborate to generate an estimate of the demand as a continuous time-varying field, and (2)~a theoretically-verified two-phase distributed optimization algorithm by which the MCAs optimize their locations to maximize the task transmission rate while ensuring that the 
assignments of tasks is proportional to the compute resources at each agent. 
We illustrate the effectiveness of our method through realistic numerical simulations, which show that our method provides up to 28\% improvement processing in throughput over a non-adaptive baseline method.

The rest of the paper is organized as follows. In Section \ref{sys_and_prob.sec}, we describe the system model and problem formulation. Our proposed method is 
presented in Section \ref{method}. Simulation results are presented in Section \ref{results}. We summarize related work in Section \ref{rel-work}. Finally, we conclude in Section \ref{conclusion}.

\section{System Model and Problem Formulation}\label{sys_and_prob.sec}

\begin{figure}
  \centering
  \includegraphics[scale=0.40]{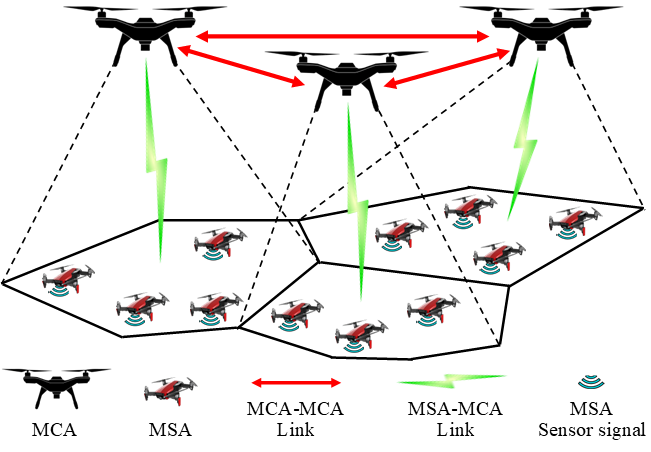}
  \caption{System model.}
  \label{fig1-model}
\end{figure}

As shown in Fig.~\ref{fig1-model}, our 
system consists of a set of MSAs that are responsible for surveilling a region of interest. The MSAs do not have the compute resources or battery power for onboard processing, so they must offload the computation tasks to more highly-resourced MCAs. The goal is to offload the MSA tasks to the MCAs and process them as quickly as possible to inform system operators and emergency 
personnel. 

To achieve this goal, the MCAs should be positioned close to MSAs so that the transmission rate is maximized. 
Since the transmission rate is inversely proportional to the transmit time, maximizing the rate also minimizes the energy expended to transmit each unit of data, which extends 
the surveillance lifetime of the MSAs. We must also ensure that the amount of work assigned to an MCA at any time does not exceed its computing capacity to avoid queuing delays at the 
MCAs. Thus, the challenge is to dynamically update the positions and task assignments of the MCAs in response to the changing demands of the MSAs (both in terms of quantity and location), 
while respecting the  capacities of the MCAs.

In the remainder of this section, we formalize our system model and the distributed task processing problem.

\subsection{System Model} \label{system-model}

There are $N$ MSAs that fly at a fixed mean sea-level (MSL) altitude of $H_{s}$ within a 3-dimensional region $\Omega$. 
Using a fixed altitude, rather than frequently ascending and descending, allows the MSAs to save energy~\cite{vfc-matching,zhou-compratemax}.
We denote the position of MSA~$n$ at time~$t$ by ${\bw_n(t) = [x_n(t), y_n(t), H_{s}]^T}$. 

The $M$ MCAs operate in the same region $\Omega$ at a fixed MSL altitude of $H_c \neq H_s$  to avoid collisions with the MSAs. 
The MCAs update their locations collaboratively to provide processing services to the MSAs. 
We denote the position of MCA~$m$ at time~$t$ by $\bu_m(t) = [x_m(t), y_m(t), H_{c}]^T$. We drop the $t$ from $\bw_n(t)$ and $\bu_m(t)$ when the context is clear.
We assume that the MCAs do not have control over, nor a priori knowledge of, the MSA trajectories.

\subsubsection{Communication Model}\label{comm-model}
We make the common assumptions that the MCAs and MSAs are equipped with omni-directional antennas and that they communicate using an Orthogonal Frequency-Division Multiple Access (OFDMA) transmission mechanism~(as in \cite{tran-jointtaskoff_multiserver,cheng-jointoff_wire_mulmec}). 
The communication channel between MCA~$m$ and MSA~$n$ is dominated by the line-of-sight link transmission with a 
distance dependent path-loss model.
The maximum transmission rate between MCA~$m$ and MSA~$n$  in bits per second (bps) at time~$t$ depends on the distance between them and is given by
\begin{align}
  r(\bu_m, \bw_n) &= B \log_2\left(1 + \frac{\beta P}{\sigma^2 \|\bu_m - \bw_n \|^2} \right)  \label{rate.eq}
\end{align}
where  $B$ is the bandwidth allocated to each MSA, $\beta$ is the channel gain at a reference distance of 1m, 
$\sigma^2$ is the power of the white Gaussian noise, and $P$ is the MSA's transmit power.

A small portion of the OFDMA subcarriers are used as control channels
for lightweight communication among MSAs and MCAs. As is common, we assume that the effects of this control channel communication on the maximum transmission rate is negligible.

\subsubsection{Workload and Offloading Model} \label{offloading_model.sec}

Each MSA~$n$ stores a queue of its unprocessed tasks. Tasks are continuously added to the queue as the MSA collects data and are removed from the queue when the MSA offloads them to an MCA.
Let $d_n(t)$~denote the quantity of unprocessed tasks at MSA~$n$ at time~$t$.
As in previous works~(cf.~\cite{li-et-al,joint-offloading,zhou-compratemax,energy-aware,uav-wpt,yan-data}), we assume that the tasks are arbitrarily divisible.
At each time~$t$, each MSA~$n$ is assigned to offload to a single MCA~$m$. 
Each MCA~$m$ has a processing capacity $c_m$, which represents the amount of tasks it can process per second (bps), i.e., its maximum throughput. 
The MSA offloads to its assigned MCA up to the rate defined in~(\ref{rate.eq}), provided the MCA has remaining capacity. If the MCA does not have remaining capacity, the MSA  stores the tasks until it is assigned to an MCA that has capacity to process them.  We note that multiple MSAs can be assigned to the same MCA simultaneously. In this case, all MSAs offload in parallel until the MCA's capacity is saturated, at which point, the MSAs queue any remaining tasks.

\subsection{Problem Formulation}

Our goal is to develop a distributed solution by which the MCAs collaboratively maximize the system throughput by maximizing the total task transmission rate, subject to the constraint that the amount of tasks assigned to each MCA~$m$ does not exceed its processing capacity $c_m$. To achieve this, at each time~$t$, the MCAs must identify both their optimal locations as well as the assignment of MSAs.
The MCAs can query the MSAs to learn the current quantities of queued tasks $d_n(t)$ and the current MSA locations $\bw_n(t)$.

We define the optimization variables $\offload{n}{t}$, $n=1 \ldots N$; the variable $\offload{n}{t}$ gives the rate at which MSA~$n$ transmits its data at time~$t$.
We also define the set of binary assignment variables $a_{m,n}(t)$, $n=1 \ldots N, m=1 \ldots M$, where $a_{m,n}(t) = 1$ if MSA~$n$ is assigned to MCA~$m$ at time~$t$, and is 0 otherwise. Let $\Omega_c$ denote the two-dimensional plane in~$\Omega$ at MSL height~$H_c$.
We formalize our problem as follows:
\begin{align}
&\underset{{\substack{\bu_m \in \Omega_c, m=1\ldots M\\ a_{m,n}, m=1\ldots M, n=1\ldots N}}}{\text{maximize}}~\sum_{m=1}^M \sum_{n=1}^N a_{m,n}(t) \offload{n}{t} \label{totaltasks.obj} \\
&\text{subject to~} \nonumber \\
&\sum_{n=1}^{N} a_{m,n}(t) \offload{n}{t} \leq c_m,~~{m=1 \ldots M} \label{capacity.con}\\
&\offload{n}{t} \leq d_n(t),~~{n=1 \ldots N} \label{demand.con}  \\
&a_{m,n}(t) \offload{n}{t} \leq r(\bu_m, \bw_n),~m=1\ldots M, n=1 \ldots N   \label{rate.con} \\
&\sum_{m=1}^M a_{m,n}(t) = 1,~~ n=1 \ldots N \label{assign.con} \\
&a_{m,n}(t) \in \{ 0, 1\},~~ m=1 \ldots M, n=1\ldots N.   \label{binary.con}
\end{align}
Here, the objective (\ref{totaltasks.obj}) is to maximize the total transmission rate.  The constraints~(\ref{capacity.con}) ensure that the total transmission rate to each MCA does not exceed its processing capacity. The constraints~(\ref{demand.con}) require that an MSA cannot offload more data than it has in its queue, and the constraints (\ref{rate.con}) require that the transmission rate between an MSA and an MCA does not exceed the maximum transmission rate defined in (\ref{rate.eq}). Finally, the constraints (\ref{assign.con}) and (\ref{binary.con}) ensure that each MSA is assigned to exactly one MCA. 

The above optimization problem presents multiple challenges. First the trajectories and the workloads of the MSAs are not known a priori, nor can they be controlled by the~MCAs. Second, even if these trajectories and locations were known, this optimization problem is a mixed-integer nonlinear programming problem, which is NP-Hard, in general. Further, both the objective and constraints are coupled across the MCA locations and assignments. Thus, even if the problem were tractable, it is not straightforward to devise a distributed solution for it.

To address these challenges, we model the workload as a continuous spatial field, a \emph{task field}, rather than considering the discrete locations of the MSAs.  
This approach allows us to devise a distributed solution by which the MCAs predict the evolution of the task field and adaptively position themselves to maximize the transmission rates while maintaining their processing capacity constraints. We describe our continuum approach and proposed solution in the next section.

\section{Continuum Approach}  \label{method}

As a first step towards making the problem tractable, we develop a continuous representation of the tasks held by the MSAs.
We define a continuous task field over the sensing airspace $\Omega_s$, the two-dimensional plane in $\Omega$ at height $H_s$.
 This field is described by a continuous function $\rho$ where ${\rho({\bx}, t) \in \mathbb{R}_{\geq 0}}$ quantifies the amount of tasks, in bits, that needs to be offloaded for processing from location $\bx \in \Omega_s$ at time~$t$.

Let $\bu(t) = [ \bu_1(t)^T, \ldots, \bu_M(t)^T]^T$ denote the positions of the MCAs at time~$t$. 
To maximize the total transmission rate, the tasks $\rho(\bx, t)$ should be assigned to the MCA~$m$ with maximal transmission rate $r(\bu_m, \bx)$.
We assign a cost to offloading one bit from location $\bx$ to $\by$, defined by
\begin{align}
  \lat{\bx}{\by} &= \frac{1}{  r( \bx , \by) } \label{latency.eq}
\end{align}
and note that minimizing $\lat{\bx}{\by}$ will maximize $r(\bx, \by)$.

We create a Voronoi partitioning of $\Omega_s$ based on the locations of the MCAs. 
The Voronoi region of MCA~$m$ is defined as 
\begin{align}
    \cV_m(\bu(t)) = \{ \bx \in \Omega_s~|~\lat{\bu_m(t)}{\bx} \leq \lat{\bu_k(t)}{\bx} \nonumber \\
 ~~~~~~~~~~~~~~~   k=1 \ldots M, k \neq m\}. 
\end{align}
Note that the Voronoi region of MCA~$m$ depends on the positions of all MCAs.
We denote the Voronoi partitioning of the entire region by:
\begin{align}
    \cV(\bu(t)) = \{\cV_1(\bu_1(t)),\ldots,\cV_M(\bu_m(t)) \}. \label{partitioning.eq}
\end{align}

For a given Voronoi partitioning, the amount of tasks assigned to MCA~$m$ at time~$t$ is given by the volume of its Voronoi region:
\begin{align}\label{nb_tasks}
  |\cV_m(\bu_m(t))| &= \int_{\cV_m} \rho (\bx, t)~\der \bx. 
\end{align}
If $|\cV_m| \leq c_m$, then the amount of tasks assigned to MCA~$m$ is less than or equal to its processing capacity.
The total cost of the tasks assigned to MCA~$m$  is:
\begin{align}
    f_m(\bu(t)) = \int_{\cV_m} \rho(\bx, t) \lat{\bu_m(t)}{\bx} \der \bx \label{m_rate.eq}
\end{align}
and the total system cost is:
\begin{align}
F(\bu) = \sum_{m=1}^M f_m(\bu(t)). \label{total_rate.eq}
\end{align}

We now reformulate problem (\ref{totaltasks.obj}) -- (\ref{binary.con}) using this continuum task field model and the Voronoi partitioning. 
\begin{align}
    \underset{\bu_m \in \Omega_c , m=1 \ldots M}{\text{minimize}}~~& F(\bu(t)) \label{opt1.prob} \\
    \text{subject to}~& |\cV_m| \leq c_m,~~~~m=1 \ldots M. \label{con1.prob}
\end{align}
The solution to this problem gives  the locations of the MCAs as well as the assignment of tasks to MCAs. By minimizing~$F$, we maximize the total  transmission rate, while the constraints (\ref{con1.prob}) ensure the capacity constraints of the MCAs are respected. Provided the Voronoi region volumes do not violate the capacity constraints, the tasks can all be offloaded and processed at their respective maximal transmission rates.

The objective (\ref{opt1.prob}) and the constraints (\ref{con1.prob}) are non-convex, and as far as we are aware, there is no single distributed algorithm to solve such a problem. Thus, we decompose the problem into tractable components that each admit a distributed solution. We then combine these components to form our full solution. We describe our proposed solution and each of its components below.

\subsection{Proposed Solution}

We propose a distributed solution to problem (\ref{opt1.prob}) - (\ref{con1.prob}).

Our solution is divided into four phases:
\begin{enumerate}
\item \textbf{Task field estimation.} In this phase, the MCAs collect information about the current state of the task queues of the MSAs and generate an estimate of $\rho$.
\item \textbf{Transmission rate maximization.} Next, the MCAs  use a distributed algorithm to determine the positions that minimize $F(\bu)$ over this estimated field $\rho$.
\item \textbf{Capacity balancing.} Then, the MCAs use a second distributed algorithm to collaboratively update these positions so that the resulting Voronoi region volumes are proportional to their processing capacities.
\item \textbf{Transmission and processing.} The MCAs move to the new positions identified in Phase 3, and the MSAs are assigned to MCAs using the Voronoi partitioning defined in (\ref{partitioning.eq}). The transmission rates are determined in a round robin fashion.  MCA~$m$ considers each MSA~$n$ one at a time and sets $\offload{n}{t}$ to the maximum of $r(\bu_m, \bw_n)$ and its remaining capacity. 
\end{enumerate}
Rather than continuously executing the first three phases, the MCAs execute them every $\window$ seconds. A smaller value of~$\window$ allows the MCAs to more accurately estimate $\rho$ and adapt their positions as it changes over time. This costs the MCAs energy to update their positions, which decreases the energy available for task processing.
 A larger value of $\window$ leads to less accurate estimates of $\rho$ and potentially lower transmission rates, but it saves in the motion energy.
 We explore different values of $\window$ in the experiments. 
 
For the first $\window$ seconds, the MCAs hover in their initial positions. After $\window$ seconds, they query the MSAs, which respond with the number of tasks they each generated in the first $\window$ seconds. The MCAs use these values as~$d_n(t)$ in phase~1. The MCAs execute phases~2 and 3 to determine their next locations, and then update their positions accordingly by flying along straight paths to their destinations. They hover in these positions until~$2\Delta$~seconds total have elapsed, and the entire process is repeated. Tasks are continuously assigned and processed as described in phase 4 using the Voronoi partitioning corresponding to the current MCA positions.

Next, we present the details of the first three phases.

\subsection{Task Field Estimation}

We consider $\rho(x)$ as a spatial Gaussian process, $GP(\mu, {\mathcal K}(\bx,\by))$ for $\bx, \by \in \Omega_s$. The constant $\mu$ is the mean of the Gaussian process,
and  ${\mathcal K}$ is the kernel function that gives the spatial correlation of $\rho$. We use the squared exponential Gaussian kernel
\begin{align}\label{kern}
  {\mathcal K}(\bx, \by) = \varphi^2 e^{\big( {-\frac{{\|\bx-\by \|}^2}{L^2}}\big)} \hfill
\end{align}
where $\varphi^2$ is the variance of the Gaussian process and $A_1$ is the length scale. Gaussian processes are used in many applications to model spatial processes~\cite{schulz2018tutorial}.

At the beginning of each window, at time $t_0$, each MCA collects the value $d_n(t_0)$ and location $\bw_n(t_0)$ from each MSA in its Voronoi region. The MCAs share these values with each other over the control channels. Then, independently, each MCA uses the observed values of $d_n(t_0)$ and $\bw_n(t_0)$, ${n=1 \ldots N}$, to construct an estimate of $\rho$ using Gaussian process regression, where $\mu$, $\varphi^2$, and $L$ are estimated from the observed values.

\subsection{Transmission Rate Maximization Algorithm}

In this phase, the MCAs collaborate to solve the unconstrained optimization problem:
\begin{align}
 \underset{\bu_m \in \Omega_c, m=1,\ldots, M}{\text{minimize}}~~& F(\bu) = \sum_{m=1}^M f_m(\bu). \label{centroidal.prob}
 \end{align}
 By doing so, the MCAs identify the positions and Voronoi partitioning that maximizes the total \emph{maximal} transmission rate. However, this partitioning does not necessarily respect the capacities of the MCAs.
  
 Each MCA communicates with nearby MCAs to solve this problem.
We say that two MCAs are \emph{neighbors} for window $b$ if their Voronoi regions are adjacent in window $b-1$. 
We define the communication graph for window $b$ as~${{\cal G}_b = (V, E_b)}$, where $V$ is the set of MCAs, and $(m,k) \in E_b$ if and only if MCAs $m$ and $k$ are neighbors in window $b-1$. 
We denote the set of neighbors of MCA~$m$ in window $b$ by ${\cal N}_b(m)$.

Each MCA~$m$ stores an estimate of the positions of all $M$~MCAs.
We let $\bud{m}_k$ denote MCA~$m$'s estimate of the position of MCA $k$,  and $\bud{m}$ denotes the concatenation of all the estimates held by MCA~$m$, i.e.,
\[ 
\bud{m} = [(\bud{m}_1)^T, \ldots, (\bud{m}_M)^T]^T. 
\]
We can then rewrite (\ref{centroidal.prob}) in the following decomposable form:
\begin{align}
 \underset{\bud{1}, \ldots, \bud{m}}{\text{minimize}}~& \hat{F}([(\bud{1})^T, \ldots, (\bud{M})^T]^T) = \sum_{m=1}^M  f_m(\bud{m}) \label{decomp1.prob} \\
\text{subject to~} & \bud{m} = \bud{k},~\text{for all } (m,k) \in E_b. \label{decomp2.prob}
\end{align}

 \begin{algorithm}
 \caption{Transmission rate maximization algorithm.}\label{alg1}
 \begin{algorithmic}[1]
 \State Initialize: $\bud{m}(0)$ to current MCA positions 
\For{$\ell=0, 1, 2, \ldots, T_1$}
\For{each MCA~$m$}
\State $\bzd{m}(\ell) \gets \proj{\bud{m}(\ell) - \eta_{\ell} \nabla f_m(\bud{m}(\ell))}$ 
\State $\bud{m}(\ell + 1) \gets \xi \sum_{j \in \mathcal{N}_b(m)} \bzd{j}(\ell)  $
\State $~~~~~~~~~~~~~~~~~~~~~~~~~~~~+ (1 - \xi |{\cal N}_b(m)|) \bzd{m}(\ell)$
\EndFor
\EndFor
\end{algorithmic}
 \end{algorithm}

The MCAs use a distributed projected gradient algorithm to solve this problem~\cite{bianchi2012convergence,zeng-nonconvex-dgd}.
We now describe the details of this algorithm. The pseudocode is shown in Alg.~\ref{alg1}. The algorithm executes in discrete iterations $\ell=0, 1, 2, \ldots, T_1$. 
Each MCA initializes its estimate $\bud{m}$ to be the current MCA positions, i.e., their positions at time $t_0$. 
In each iteration~$\ell$, each MCA~$m$ first computes the derivative of $f_m$ with respect to its estimate $\bud{m}$. It then performs a gradient step on this estimate with step size $\eta_{\ell}$ and projects the result of this gradient step onto $\Omega_c$, where $\proj{\bx}$ denotes the Euclidean projection onto the region $\Omega_c$.
To complete the iteration, the MCA updates its estimate by taking a weighted average of its own estimate and those of its neighbors; MCA~$m$ applies weight~$\xi$ to each of its neighbor's estimates and weight~$(1 - \xi |{\cal N}_b(m)|)$ to its own estimate. 

The MCAs repeat this iteration until a desired convergence is achieved (typically within a hundred iterations in our experiments).
Intuitively, through the projected gradient steps of Alg.~\ref{alg1},  each MCA updates its estimates of the positions to maximize the transmission rate  in its own Voronoi region, while the averaging steps drive the estimates to agreement. 

We now summarize the convergence behavior of Alg.~\ref{alg1}.
Define the set of stationary points on $(\Omega_c)^{M^2}$ as 
${\mathcal{L} = \{ \bx \in (\Omega_c)^{M^2}~|~ \nabla \hat{F}(\bx) \in \mathcal{C}(\bx)}\}$,
where $\mathcal{C}(\bx)$ is the normal cone, i.e., ${\mathcal{C}(\bx)= \{ \bv \in \mR^{2M}~|~ \forall \bu' \in (\Omega_c)^{M^2}, 
\bv^T(\bx - \bu')' \geq 0\}}$.
With this definition, we present the following theorem.
\begin{theorem}\label{alg1.thm}
Let $\Omega_c$ be a nonempty compact convex set,  and assume that $\hat{F}(\mathcal{L})$ has an empty interior and that $\xi \leq \frac{1}{\delta}$, where $\delta$ is the maximum vertex degree of ${\cal G}_b$\footnote{If ${\cal G}_b$ is bipartite, we require $\xi < \frac{1}{\delta}$.}.
Let the step size $\{\eta_{\ell}\}_{\ell \geq 0}$ be such that $\sum_{\ell=0}^{\infty} \eta_{\ell} = \infty$ and $\sum_{\ell=0}^{\infty} \eta_\ell^2 < \infty$.
Then, the sequence $\{[\bud{1}(\ell)^T, \ldots, \bud{M}(\ell)^T]^T\}_{\ell \geq 0}$ converges to the set ${\{ \mathbf{1} \otimes \bx~|~ \bx \in \mathcal{L}\}}$\footnote{The symbol $\otimes$ denotes the Kronecker product.}. Moreover, 
$\frac{1}{M} \sum_{m=1}^M \bud{m}(\ell)$ converges to a connected component of $\mathcal{L}$. 
\end{theorem}
The proof of this theorem follows from Theorem 1 in \cite{bianchi2012convergence} and the fact that problem (\ref{decomp1.prob})-(\ref{decomp2.prob}) satisfies certain technical conditions. The full proof is given in the appendix. 

Theorem~\ref{alg1.thm} shows that Alg.~\ref{alg1} converges to a stationary point of the transmission rate maximization problem, and further, that the MCAs position vectors $\bud{m}$, $m=1 \ldots M$, converge to agreement at this fixed point.

 \begin{algorithm}
 \caption{Capacity balancing algorithm.}\label{alg2}
  \begin{algorithmic}[1]
 \State Initialize: $\bu_m(0)$ to result of Alg.~\ref{alg1} 
 \For{$\ell=0, 1, 2, \ldots, T_2$}
\For{each MCA~$m$}
\State $\bd_m(\ell) \gets \sum_{j \in {\cal N}_b(m)} \left( \left(\frac{\int_{\cV_m} \rho(\bx) d \bx}{c_m} - \frac{\int_{V_j} \rho(\bx) d \bx}{c_j}\right) \times \right.$
\State $~~~~~~~~~~~~~~~~~~~~\left. n_{mj} \int_{\cV_m \cap \cV_j} \rho(\gamma) d \gamma  \right)$
\State $\bu_m(\ell+1) = \bu_m(\ell)- \alpha \bd_m(\ell)$
 \EndFor
\EndFor
\end{algorithmic}
 \end{algorithm}

\subsection{Capacity Balancing Algorithm}
While Alg.~\ref{alg1} aims to optimize the Voronoi partitioning for transmission rate, it may result in some Voronoi regions with workloads that exceed the capacities of their MCAs. To optimize the system throughput, the MCAs  must  have the capacity to process their assigned tasks. We next present a distributed algorithm to adjust the MCA positions and thus adjust the Voronoi partitioning, so that the region volumes are proportional to the processing capacities of the MCAs. 

We define the functions $g_m(\bu) = |\cV_m|^2$, $m=1 \ldots M$, and the function $G$:
\begin{align}
G(\bu) = \sum_{m=1}^M \frac{1}{c_m} g_m(\bu).
\end{align}
We define the optimization problem:
\begin{align}
 \underset{\bu_m \in \Omega_c, m=1 \ldots M}{\text{minimize}}~~ G(\bu). \label{capacity.prob}
\end{align}
As shown in~\cite{5609192}, the solution to problem (\ref{capacity.prob}) creates a Voronoi partitioning where the total workload assigned to each MCA is proportional to its capacity.
This is formalized in the following lemma.
\begin{lemma}[Lemma 1 in~\cite{5609192}]
Let $x_m$, $m=1 \ldots M$ be real variables, subject to the constraint that $\sum_{m=1}^M x_m = X$ where $X$ is a constant. The function $\sum_{m=1}^M(x_m^2/c_m)$, with constants~$c_m$, attains its minimum when $x_m/c_m = X/\sum_{m=1}^M c_m$, for $i=1 \ldots M$.
\end{lemma}

We present a discrete-time distributed gradient algorithm, based on the continuous-time solution presented in~\cite{5609192}, to solve  problem~(\ref{capacity.prob}).
The pseudocode is shown in Alg.~\ref{alg2}.
The MCAs initialize their positions to the positions determined by the transmission rate maximization algorithm. 
In each iteration~$\ell$, each MCA communicates with its neighbors in the graph~${ \cal G}_b$ using the control channels to exchange their values $\bu_m(\ell)$. Each MCA~$m$ then updates $\bu_m(\ell+1)$ to be a weighted average of its $\bu_m(\ell)$ and that of its neighbor MCAs:
\begin{align}
\bu_m(\ell+1) &= \bu_m(\ell)- \alpha \bd_m(\ell)
\end{align}
where $\alpha$ is the step size, and
\begin{align}
\bd_m(\ell) &= \sum_{j \in {\cal N}_b(m)} \left( \left(\frac{\int_{\cV_m} \rho(\bx) d \bx}{c_m} - \frac{\int_{V_j} \rho(\bx) d \bx}{c_j}\right) \times \nonumber  \right. \\
 &~~~~~~~~~~~~~~~~~~\left. n_{mj} \int_{\cV_m \cap \cV_j} \rho(\gamma ) d \gamma \right). \label{direction.eq}
\end{align}
The weights used in the averaging depend on the volumes of the Voronoi regions. Here, $n_{mj}$ denotes the unit normal of the shared boundary of the Voronoi regions of MCAs $m$ and $j$. Each term in the sum in (\ref{direction.eq}) exerts a ``push'' or ``pull'' on MCA~$m$'s position proportional to how balanced $m$'s capacity is with respect to its neighbor.
The MCAs execute this algorithm until it converges to a balanced partitioning (typically within a hundred iterations in our experiments).
The convergence behavior of Alg.~\ref{alg2} is given in the following theorem.
\begin{theorem}\label{alg2.thm}
For an appropriately chosen step size $\alpha$, Alg.~\ref{alg2} converges asymptotically to an equilibrium where the MCA positions are such that the volume of each Voronoi region is proportional to the MCA's capacity. 
\end{theorem}
Theorem~\ref{alg2.thm} follows directly from Theorem 1 in~\cite{5609192}.

By initializing Alg.~\ref{alg2} with positions that optimize for transmission rate, and then using Alg.~\ref{alg2} to adjust the positions to balance the workload to match the capacities, the resulting Voronoi partitioning incorporates both transmission rate maximization and processing capacity to optimize the total system throughput.

\section{Simulation Results}\label{results}
We evaluate our collaborative task processing method through numerical simulations. We first detail our simulation setup, followed by experimental results on the task field estimation and distributed task processing.

\begin{table}[htbp]
  \caption{Simulation parameters.} \label{tab:tab1}
  \centering
  \begin{tabular}{l l ||l l}
  \textbf{Parameters} & \textbf{Value} & \textbf{Parameter} & \textbf{Value} \\
    \hline\hline 
    Region size & $(5000~\text{m})^2$ & Sub-region size & $(100~\text{m})^2$ \\
    \hline
    $H_c$ & $100$ m & $H_s$ & $50$ m \\
    \hline
    MCA speed & $25$ m/s & MSA speed & $10 - 20$ m/s \\
    \hline
   Sub-region duration & $50 - 60$ s & Sub-region pause & $2 - 5$ s \\
    \hline \hline
    $B$ & $0.2$ Mbs & $P$ & $40$ dBm \\
    \hline
    $\beta$ & $-50$ dB & $\sigma^2$ & $-60$ dBm \\
    \hline \hline
    $T_1$ & $100$ & $\eta_{\ell}$ & $1e^{-4}$ \\ 
    \hline
    $T_2$ & $200$ & $\alpha$ & $1e^{-2}$ \\
    \hline
    $\xi$ & $1/\delta$ \\
    \cline{1-2}
    \end{tabular}
\end{table}

\subsection{Simulation Setup}

We implemented our simulations using Matlab. To simulate continuous time, we discretize it into $0.1$s time steps. For Gaussian process regression, we use the Matlab Statistics and Machine Learning toolbox.

We consider a scenario with $N=50$ MSAs that operate in a $(5000~\text{m})^2$ area at a fixed MSL altitude of $H_s = 50$~m.
The MSAs surveil the region using a modified random waypoint model based on~\cite{rwp-model}. Each MSA starts at a randomly initialized location in the region and chooses a destination at random within the region. The MSA flies to this destination along the shortest path and then flies within a bounded sub-region of $(100~\text{m})^2$ using a random waypoint trajectory for a period of time chosen uniformly at random between 50s and 60s. The MSA pauses for a random time between 2s and 5s, and repeats the entire process. The speed of each MSA, in m/s, is randomly selected from the interval $[10, 20]$. 

As the MSAs fly, they collect sensing data. Rather than collecting data at a constant rate, the collection rate increases as the MSAs are close to points of interests, e.g., representing wildfire fronts.
For the sake of comparison, we standardize the total number of tasks generated per second to $6\times10^6$~bps, and each MSA generates a number of tasks proportional to its distance from the point(s) of interest.
 We consider two scenarios: 
\begin{enumerate}
\item \textbf{Fixed points of interest.} We use two points of interest at locations $\mathbf{p}_1$ and $\mathbf{p}_2$. The point locations $\mathbf{p}_1$ and $\mathbf{p}_2$ are selected uniformly at random and remained fixed for the duration of the simulation. The quantity of tasks generated at MSA~$n$ at time~$t$ is proportional to~$\| \bw_n(t) - \mathbf{p}_1\|^{-1.5} + \| \bw_n(t) - \mathbf{p}_2\|^{-1.5}$.
\item \textbf{Moving point of interest.} We use a single point of interest, initially at location ${\mathbf{p}(0) = [4500,4500,0]^T}$, which  moves $500$m every $15$s at an angle of $-45^{\circ}$ across the region.  The quantity of tasks generated at MSA~$n$ at time~$t$ is proportional to~$\| \bw_n(t) - \mathbf{p}(t)\|^{-1.5}$.
\end{enumerate}

For the task processing experiments, there are ${M=6}$~MCAs. The MCAs fly at fixed MSL altitude of $H_c = 100$~m. At the beginning of each window, starting with window 2, the MCAs pause for 0.1s to execute our algorithm and determine their new positions. 
They then fly to their new positions at a constant speed of $25$~m/s and hover at their new positions until the end of the time window. The simulation parameters are given in Table~\ref{tab:tab1}.

\subsection{Task Field Estimation} \label{estex.sec}

We first explore the accuracy of the task field estimation phase.  The MSAs collect data for time window $(0, \Delta]$. We then sample $d_n(\Delta)$, the amount of tasks they collected over these $\Delta$ seconds, from all 50 MSAs. We use these observations to generate the estimate $\rho$ via Gaussian process regression.
In our solution, this estimate of $\rho$ is used to determine the MCA positions for the time window $(\Delta, 2\Delta]$. We therefore compare the estimate $\rho$ against the distribution of tasks collected in time window $(\Delta, 2\Delta]$. Since we are interested in studying the estimation accuracy for a single time window, we use fixed points of interest in these experiments.

For the purposes of comparison, we discretize the region into $(50~\text{m})^2$ cells. For each MSA~$n$, we consider the cells that it visits in window $(\Delta, 2\Delta]$, and we assign an equal portion of $d_n(t)$ to each of these cells. The resulting discrete field~${\mathbf{T} \in \mR^{100 \times 100}}$ is the \emph{discretized task field} for window~$(\Delta, 2\Delta]$. We generate the discretized task field estimate $\mathbf{E} \in \mR^{100 \times 100}$ by computing the value of $\rho$ at the center of each cell.  To compare $\mathbf{T}$ and $\mathbf{E}$, we use the normalized mean-square error (NMSE) ${\| \mathbf{E} - \mathbf{T} \|_F / (\text{max}(\mathbf{T}) - \text{min}(\mathbf{T}))}$. Here, $\| \cdot \|_F$ denotes the Frobenius norm.

\begin{figure}
    \centering
    \includegraphics[scale=.23]{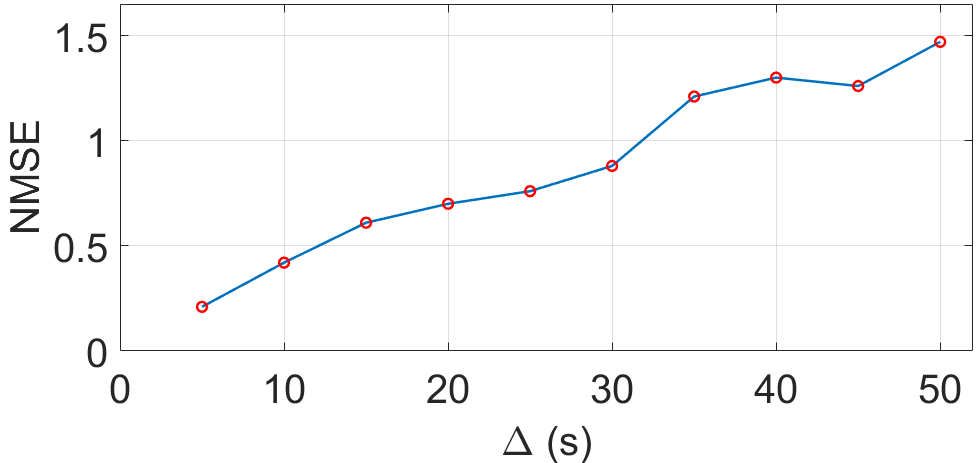}
    \caption{Change in NMSE with increasing $\Delta$.} \label{estacc.fig}
\end{figure}
Fig.~\ref{estacc.fig} shows the NMSE for various values of $\Delta$. For each~$\Delta$, the NMSE is the average of three experiments. As can be seen, the NMSE increases as~$\Delta$
increases. This is as anticipated since the smaller the value of $\Delta$, the closer our approximation is to a continuous time model. 

\begin{figure*}
\centering
    \begin{subfigure}{0.23\textwidth}
        \includegraphics[width=\linewidth]{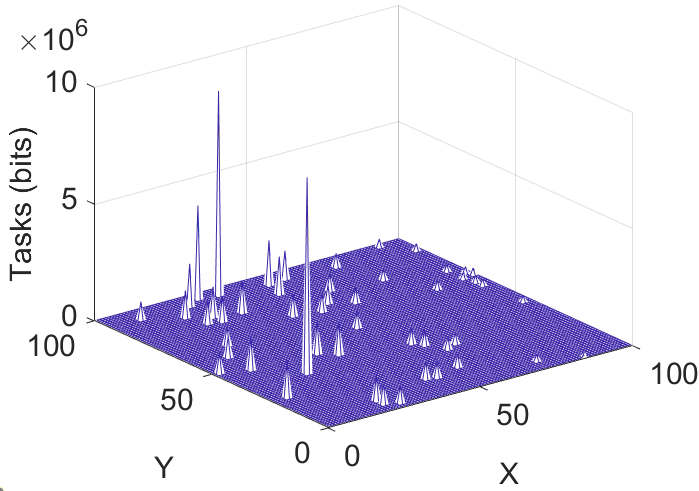}
        \caption{Observations for $\Delta=10$s.} \label{S10.fig}
    \end{subfigure} 
    \begin{subfigure}{0.23\textwidth}
        \includegraphics[width=\linewidth]{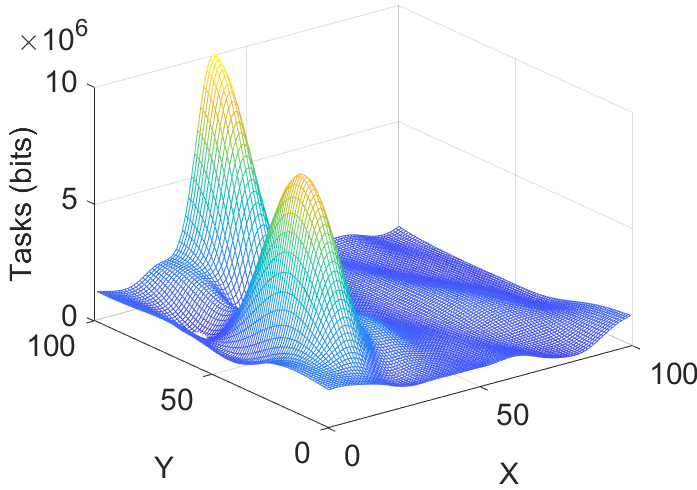}
        \caption{Estimated field for $\Delta=10$s.} \label{E10.fig}
    \end{subfigure}     
    \begin{subfigure}{0.23\textwidth}
        \includegraphics[width=\linewidth]{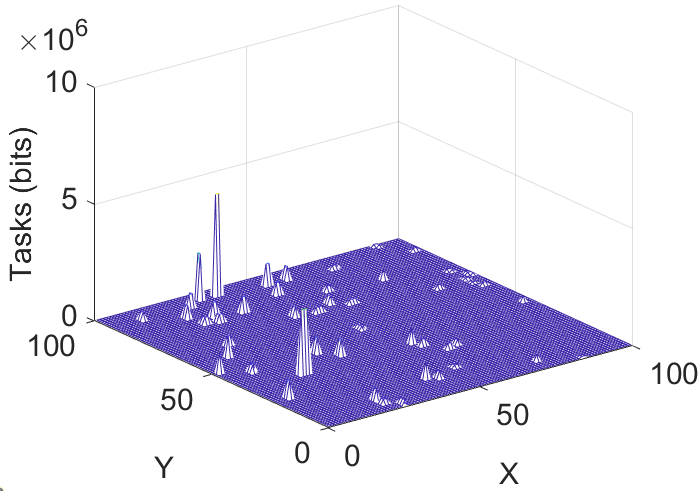}
        \caption{Task field for $\Delta=10$s.} \label{W10.fig}
    \end{subfigure}\\
    \begin{subfigure}{0.23\textwidth}
        \includegraphics[width=\textwidth]{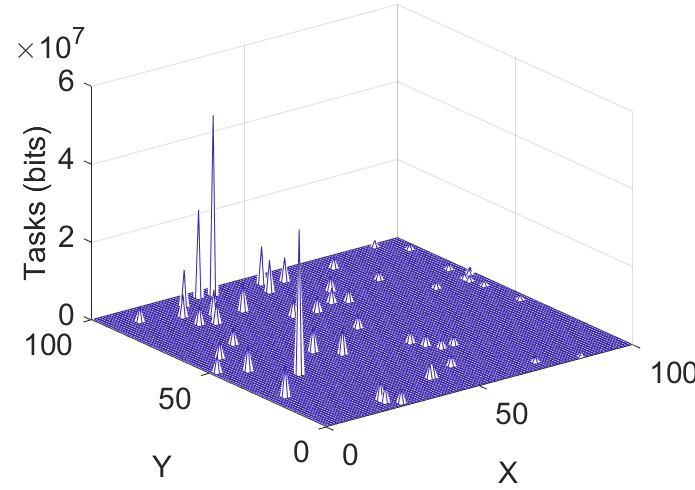}
        \caption{Observations for  $\Delta=50$s.} \label{S50.fig}
    \end{subfigure}   
    \begin{subfigure}{0.23\textwidth}
        \includegraphics[width=\textwidth]{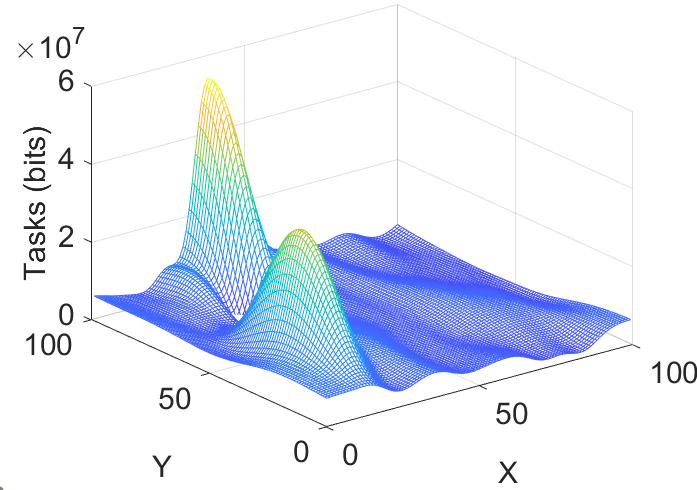}
        \caption{Estimated field for ${\Delta=50\text{s}}$.} \label{E50.fig}
    \end{subfigure}
    \begin{subfigure}{0.23\textwidth}
        \includegraphics[width=\textwidth]{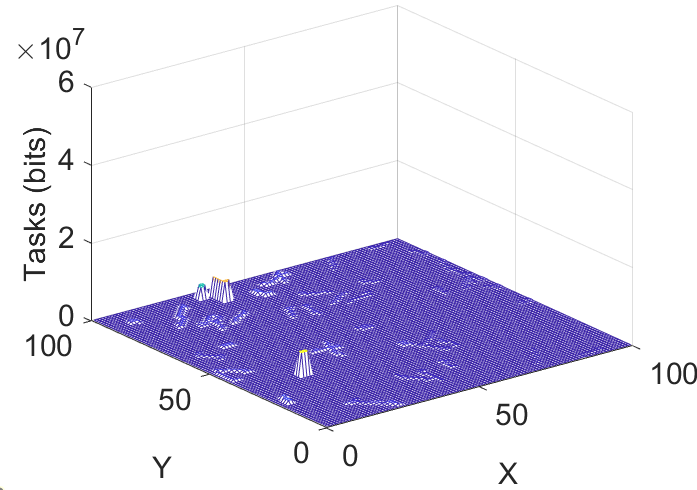}
        \caption{Task field for $\Delta=50$s.} \label{W50.fig}
    \end{subfigure}

    \caption{Examples of the observed task queues, estimated task fields from these observations $\mathbf{E}$ , and the discretized task fields $\mathbf{T}$ for the window,  for $\Delta=10$s and $\Delta=50$s.}
    \label{estex.fig}
\end{figure*}

This trend is further corroborated in Fig.~\ref{estex.fig}. In Figs.~\ref{S10.fig} and \ref{S50.fig}, we show the observed values $d_n(\Delta)$, each in the cell where its corresponding MSA~$n$ was located at time $\Delta$, 
for ${\Delta=10\text{s}}$ and ${\Delta=50\text{s}}$, respectively. Figs.~\ref{E10.fig} and \ref{E50.fig} show the corresponding estimates of the task fields. We can see that for both values of $\Delta$, the estimate captures the locations and magnitudes of the observations.
The estimated field is smooth; the benefit of this smoothing is that it can, in some part, account for the movement of the MSAs throughout the subsequent window.  Figs.~\ref{W10.fig} and \ref{W50.fig} show the discretized task fields for the subsequent window. We observe that the estimates reflect the locations of higher volumes of tasks, however, the magnitudes of the estimate are much larger than those of the discretized task fields. This is due to the fact that in our discretized task field, the MSA's tasks are split over all cells that it visits in the window. The magnitude discrepancy is larger for $\Delta=50$s than $\Delta=10$s because the MSAs move for a longer period of time and thus cover more cells. These results indicate that while the estimates differ in magnitude from the discretized task fields, they are still valuable for selecting the locations for MCAs to optimize the throughput.

\subsection{Collaborative Task Processing}

We next explore the performance of our full proposed solution. Based on the results of the previous section, we select two smaller values of $\Delta$, $\Delta=10$s and $\Delta = 20$s.  Each experiment lasts $120$s.

We evaluate three approaches:
\begin{enumerate}
\item \textbf{Baseline.} The MCA positions remained fixed for the duration of the experiment. 
\item \textbf{Transmission rate maximization.} Every $\Delta$ seconds, the MCAs generate an estimate of $\rho$ and then execute the transmission rate maximization algorithm and update their positions accordingly.
\item \textbf{Full solution.} Every $\Delta$ seconds, the MCAs generate an estimate of $\rho$ and then execute the transmission rate maximization algorithm followed by the capacity balancing algorithm. They then update their positions accordingly.
\end{enumerate}
For all three approaches, the MCAs are initially positioned in two rows of three, so that each MCA has the same size Voronoi region.
For the baseline approach, each MSA continuously offloads to its closest MCA, provided that the MCA has capacity to process the workload. If the closest MCA does not have capacity, the MSA  queues its data until the closest MCA has available capacity. Note that since the MSA moves throughout the experiment, the closest MCA may be different at different times.
For the transmission rate maximization and full solution approaches, the MCAs start executing the algorithms after the first $\Delta$ seconds.
At the start of each subsequent window, the MCAs pause for $0.1$s to generate $\rho$ and compute their new positions. They then move to those new positions.
The MSAs continuously offload to the MCAs assigned to them by the Voronoi partitioning. 
If an MCA does not have sufficient capacity, the MSA queues any remaining tasks until it is assigned to an MCA that does have capacity as in the baseline approach. 

For ease of comparison, we set the total processing capacity of the system to be $6\times10^6$~bps so that it matches the total task generation rate. We study two scenarios, a \emph{homogenous capacity} scenario, where each MCA has the same capacity of $10^6$~bps, and a \emph{heterogeneous capacity} scenario, where MCA~1 has capacity~$2 \times 10^6$~bps, MCAs 2, 3, and 4 have capacities~$10^6$~bps, and MCAs 5 and 6 have capacities $0.5 \times 10^6$~bps.

For each experiment, we calculate the total amount of tasks that are processed in two ways. We show the \emph{cold start total}, where we compute the total number of tasks processed over the entire $120$s.
In all three approaches, the MCAs have the same positions, and therefore the same amount of tasks processed in the first window, and so we also calculate the \emph{warm start total}, which excludes the first $\Delta$ seconds, so as to only include the time in which the MCAs adapt their positions and task assignments. We also calculate the percentage improvement of our full solution over the baseline. All results are the averages over three experiments.

\subsubsection{Fixed points of interest} 

\begin{table}
\caption{Total tasks processed ($\times 10^8$) and warm start percent increase over baseline, with homogeneous capacities and fixed points of interest.} \label{fixedhotspotshom.tab}
\centering
\begin{tabular}{|c|c|c|c|c|c|c|}
\hline
 \multirow{2}{*}{\textbf{Approach}}&    \multicolumn{3}{|c|}{$\mathbf{\Delta = 10\text{s}}$} &  \multicolumn{3}{|c|}{$\mathbf{\Delta = 20\text{s}}$} \\
 \cline{2-7}
& \textbf{cold} &\textbf{warm} & \textbf{\% inc.} & \textbf{cold} & \textbf{warm} & \textbf{\% inc.} \\
    \hline
    Baseline & 4.54 & 4.14 & & 4.54 & 3.75 & \\
    \hline
    Rate Max. & 4.56 & 4.17 & 1 & 4.54 & 3.75 & 0 \\
    \hline
    Full Solution  & 5.29 & 4.90 & 18 & 5.01 & 4.22 & 13 \\
   \hline
\end{tabular}
\end{table}

\begin{table}
\caption{Total tasks processed ($\times 10^8$) and warm start percent increase over baseline, with heterogeneous capacities and fixed points of interest.} \label{fixedhotspotshet.tab}
\centering
\begin{tabular}{|c|c|c|c|c|c|c|}
\hline
 \multirow{2}{*}{\textbf{Approach}}&    \multicolumn{3}{|c|}{$\mathbf{\Delta = 10\text{s}}$} &  \multicolumn{3}{|c|}{$\mathbf{\Delta = 20\text{s}}$} \\
 \cline{2-7}
& \textbf{cold} &\textbf{warm} & \textbf{\% inc.} & \textbf{cold} & \textbf{warm} & \textbf{\% inc.} \\
    \hline
    Baseline & 3.88 & 3.54 & & 3.88 & 3.21 & \\
    \hline
    Rate Max. & 3.87 & 3.53 & 0 & 3.86 & 3.19 & -1 \\
    \hline
    Full Solution & 4.78 & 4.45 & 26 & 4.45 & 3.78 & 18 \\
   \hline
\end{tabular}
\end{table}

Tables~\ref{fixedhotspotshom.tab} and \ref{fixedhotspotshet.tab} show the results for the homogeneous and heterogeneous capacities, 
respectively, for the fixed points of interest scenario.
 The transmission rate maximization approach does not show much benefit over the baseline, indicating that optimizing for transmission rate alone is not sufficient to maximize the system throughput.  
We observe that the full solution yields significant improvement in total tasks processed over the baseline. Further, the benefit is greater for the smaller value of $\Delta$, with an $18\%$ improvement over the baseline for the warm start total with homogeneous capacities, and a $26\%$ improvement for heterogeneous capacities. These results show the importance of aligning the task assignment with the MCA capacities---a novel feature of our proposed solution. We also observe that in all cases, the total amount of tasks processed is less than the total amount of tasks generated ($7.2 \times 10^8$~bits for cold start; $6.6 \times 10^8$~bits and $6.0 \times 10^8$~bits for warm start for $\Delta = 10$s and $20$s, respectively). We believe this is due to two factors. First, the MCAs spend part of each window moving to their new positions. During this transition period, the Voronoi partitioning is sub-optimal with respect to transmission rates and capacities. Second, any inaccuracies in the estimate of $\rho$ lead the MCAs to select positions that are less compatible with the true distribution of tasks. Thus, there is opportunity to improve on our solution with more sophisticated task field estimation techniques. 

\begin{figure*}
\centering
    \begin{subfigure}{0.23\textwidth}
        \includegraphics[width=\linewidth]{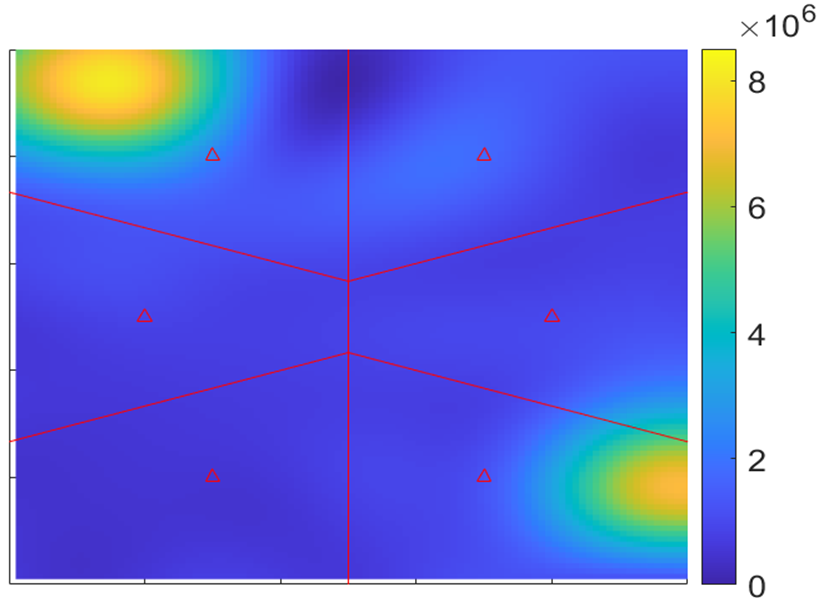}
        \caption{Window 1.}
    \end{subfigure}
    \begin{subfigure}{0.23\textwidth}
        \includegraphics[width=\linewidth]{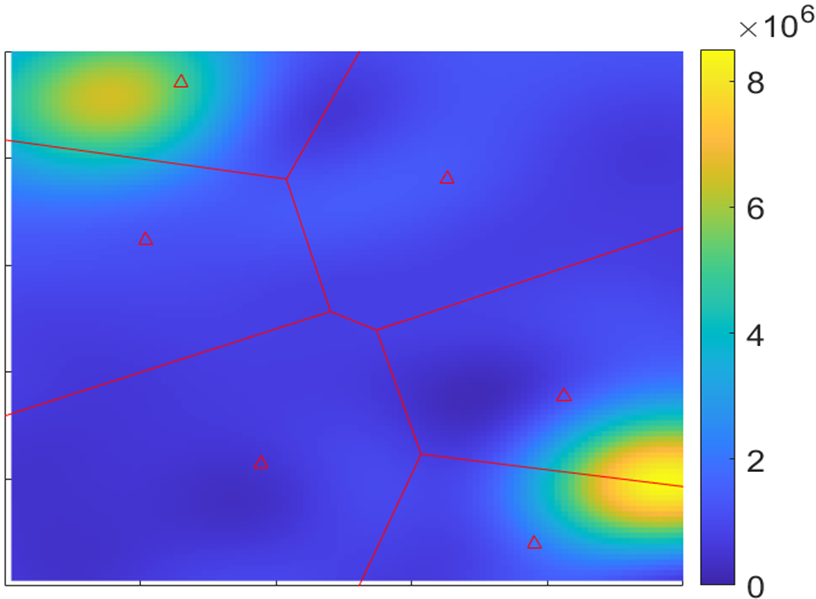}
        \caption{Window 4.}
    \end{subfigure}
    \begin{subfigure}{0.23\textwidth}
        \includegraphics[width=\linewidth]{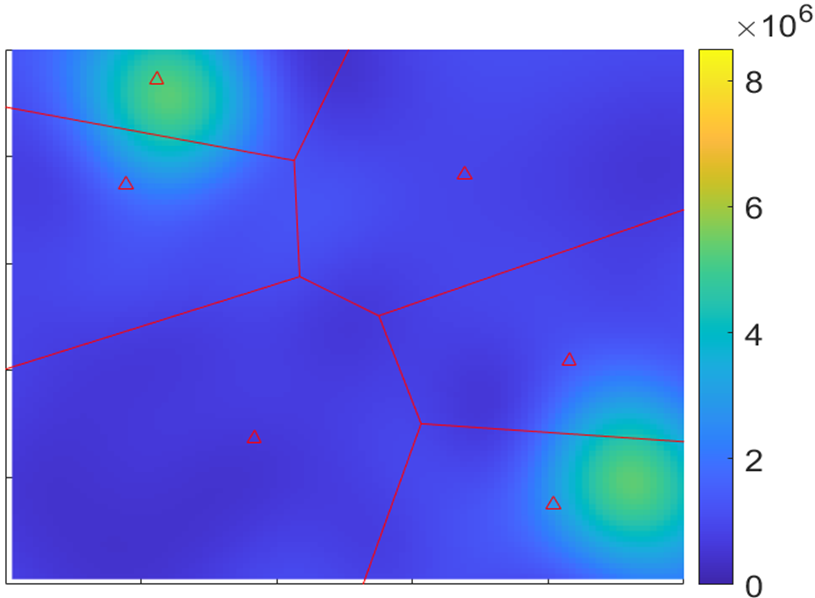}
        \caption{Window 8.}
    \end{subfigure}
    \caption{Examples of homogeneous MCAs adjusting their positions to the task field for fixed points of interest for $\Delta=10$s. The triangles represent the MCAs. The background color represents the intensity of the estimated task field.} \label{fixedex.fig}
\end{figure*}

Fig.~\ref{fixedex.fig} shows the evolution of the MCA locations and the Voronoi partitioning over multiple time windows for the homogeneous capacity scenario. The MCA locations and region boundaries are superimposed on a heat map of $\rho$ for that window.
 Initially (window 1), each MCA has a similar-sized Voronoi region; however, the volume of tasks in each cell is very different. In window 4, the Voronoi partitioning more accurately reflects the task distribution, and in window 8, the Voronoi partitioning is further adapted to minor changes in the tasks distribution that result from the MSAs' mobility and residual queued tasks.

\subsubsection{Moving point of interest} 

\begin{table}
\caption{Total tasks processed ($\times 10^8$) and warm start percent increase over baseline, with homogeneous capacities and moving point of interest.} \label{movhotspotshom.tab}   
\centering
\begin{tabular}{|l|c|c|c|c|c|c|}
\hline
 \multirow{2}{*}{\textbf{Approach}}&    \multicolumn{3}{|c|}{$\mathbf{\Delta = 10\text{s}}$} &  \multicolumn{3}{|c|}{$\mathbf{\Delta = 20\text{s}}$} \\
 \cline{2-7}
& \textbf{cold} &\textbf{warm} & \textbf{\% inc.} & \textbf{cold} & \textbf{warm} & \textbf{\% inc.} \\
    \hline
    Baseline & 4.09 & 3.75 & & 4.09 & 3.42 & \\
    \hline
    Rate Max.  & 4.25 & 3.91 & 4 & 4.25 & 3.58 & 5 \\
    \hline
    Full Solution  & 4.88 & 4.55 & 21 & 4.81 & 4.14 & 21 \\
   \hline
\end{tabular}
\end{table}

\begin{table}
\caption{Total tasks processed ($\times 10^8$) and warm start percent increase over baseline, with heterogeneous capacities and moving point of interest.} \label{movhotspotshet.tab}
\centering
\begin{tabular}{|c|c|c|c|c|c|c|}
\hline
\multirow{2}{*}{\textbf{Approach}}&    \multicolumn{3}{|c|}{$\mathbf{\Delta = 10\text{s}}$} &  \multicolumn{3}{|c|}{$\mathbf{\Delta = 20\text{s}}$} \\
\cline{2-7}
& \textbf{cold} &\textbf{warm} & \textbf{\% inc.} & \textbf{cold} & \textbf{warm} & \textbf{\% inc.} \\
\hline
Baseline & 3.33 & 3.06 & & 3.33 & 2.80 & \\
\hline
Rate Max.  & 3.52 & 3.25 & 6 & 3.43 & 2.90 & 3 \\
\hline
Full Solution  & 4.20 & 3.93 & 28 & 4.06 & 3.53 & 26 \\
\hline
\end{tabular}
\end{table}

Tables~\ref{movhotspotshom.tab} and \ref{movhotspotshet.tab} show the results for the homogeneous and heterogeneous capacities, 
respectively for the moving point of interest. There is a small increase in the total number of tasks processed for the transmission rate maximization approach for both capacity types. Since the task field changes as the point of interest moves, it is intuitive that adapting the MCA positions should improve the overall transmission rate, thus improving the system throughput. 
Our full solution yields significant improvement over the baseline. For MCAs with homogeneous capacities, our full solution achieved a $21\%$ improvement in the warm start total  for both $\Delta = 10$s and $\Delta=20$s. 
For the heterogeneous capacities, we observe improvements of $28\%$ and $26\%$ for $\Delta=10$s and $\Delta=20$s, respectively. 
This is a greater improvement than observed for the fixed points of interest, 
which shows a key benefit of our method in that it adapts to both small and large changes in the task field.
As in the previous set of experiments, no approach processes all of the generated tasks; we believe the same reasons apply to this scenario. 
We also observe that in most cases, our full solution performs better with $\Delta=10$s than with $\Delta=20$s. This is expected because the estimate of $\rho$ is more accurate for the smaller~$\Delta$. The improvement over $\Delta=20$s is not very large though, indicating that it is possible to save on the motion energy of the MCAs without much decay in the system throughput.

\begin{figure*}
\centering
    \begin{subfigure}{0.23\textwidth}
        \includegraphics[width=\linewidth]{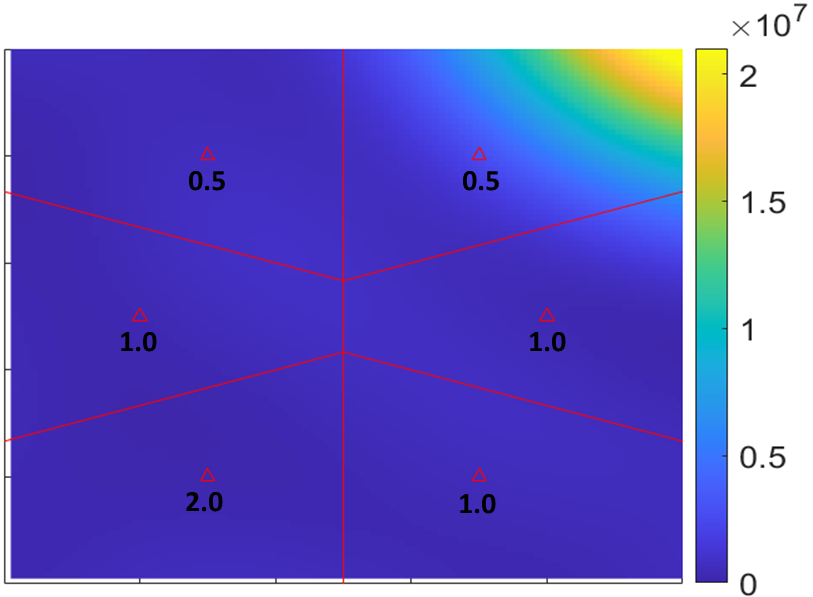}
        \caption{Window 1.}
    \end{subfigure}
    \begin{subfigure}{0.23\textwidth}
        \includegraphics[width=\linewidth]{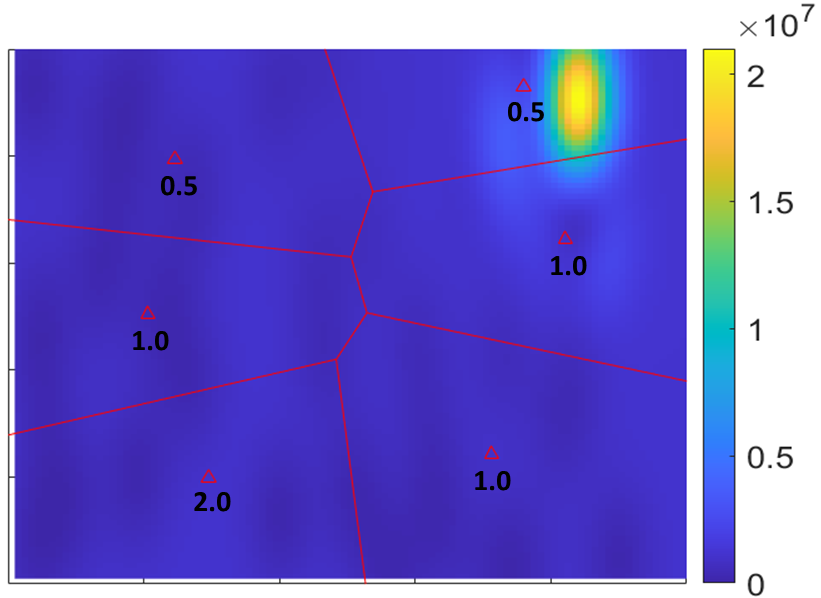}
        \caption{Window 4.}
    \end{subfigure}
    \begin{subfigure}{0.23\textwidth}
        \includegraphics[width=\linewidth]{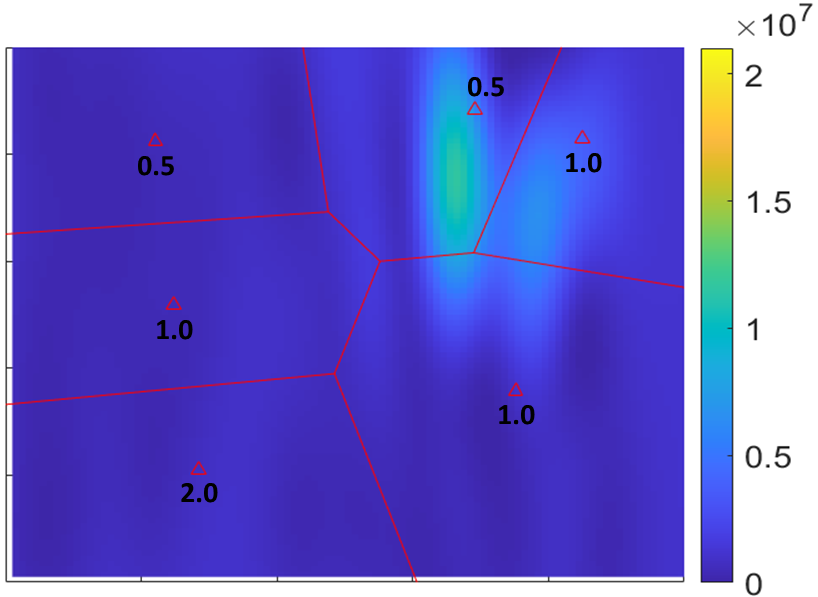}
        \caption{Window 8.}
    \end{subfigure}
    \caption{Examples of heterogeneous MCAs adjusting their positions to the task field for a moving point of interest for $\Delta=10$s. The triangles represent the MCAs. The background color represents the intensity of the estimated task field.}
    \label{moveex.fig}
\end{figure*}

Fig.~\ref{moveex.fig} shows the evolution of the MCA locations and the Voronoi partitioning over multiple time windows for the heterogeneous capacity scenario with a moving point of interest. As the point of interest moves, the task field distribution changes accordingly, and in response,
the MCAs adapt their positions and Voronoi regions. As expected, the MCA position changes are more pronounced than in the fixed points of interest setting since the changes in the task field are more significant. These figures illustrate the power of our method to autonomously adapt to changes in the task demand.

\section{Related Work}\label{rel-work}

Several works have proposed solutions using a single UAV as an MEC server for users on the ground. 
The majority of these~\cite{li-et-al,joint-offloading,zhou-compratemax,noma-6g} assume that user demand is static and known to the UAV, and they propose methods to design a UAV trajectory that minimizes the motion energy cost while also optimizing for offloading latency or rate.
\cite{energy-aware} proposes a solution to optimize the UAV trajectory when the user demand distribution changes but in a constrained and known way along straight road segments. In contrast, in our setting, the demand distribution is continuously changing, with no constraints. 
Finally, \cite{uav-wpt} proposes a solution for a single UAV system that can adapt to changes in the  user demand, however, the limitation to a single MEC UAV simplifies the problem setting over the one we study.

Several multi-UAV MEC systems for ground users have also been proposed. These works focus on designing the UAV trajectories and task assignments to optimize for similar performance measures as the single UAV systems. One line of work assumes that the user demand is fixed and known. Due to the intractability of the problem, these works utilize various relaxation approaches and meta-optimization methods such convex relaxation~\cite{wang-ofld-drl,huang-opt}, Deep Reinforcement Learning~\cite{zhao-multi-agent-drl}, and Differential Evolution~\cite{yang-load-bal}. The solutions proposed in these works are all centralized, and this centralization, along with the assumption about fixed demand, make them unsuitable for our setting.
A few works support dynamic user demand, e.g.,~\cite{yan-data}, which also proposes a solution based on Differential Evolution, \cite{liao-hotspot}, which transforms the problem to a maximum clique problem, and \cite{online-uav-mounted}, which formulates the problem as an NP-Hard bin packing problem and provides an online heuristic solution. The algorithms in these three works are centralized, which limits their applicability in disaster scenarios that do not have the infrastructure to support centralized computation. 

The recent work by Chen et al.~\cite{vfc-matching} studies the problem of task offloading from UAVs to MEC servers on ground vehicles. The work assumes that 
the trajectories of the UAVs and ground vehicles are known and cannot be altered, and the UAVs pay the ground vehicles for use of the MEC servers. They present a centralized solution based on maximal matching that optimizes for the UAV's satisfaction and the ground vehicle's payoff. The problem considered in this work is distinctly different from ours in that the MEC server trajectories cannot be controlled.

Finally, a work that is similar in spirit to ours is~\cite{dist-3D}, which
considers a setting where UAVs serve as aerial base stations for a time-varying continuous distribution of ground user demand. 
The authors present a distributed algorithm that adapts the UAV positions so as to maximize the network coverage of the ground users. 
While this work also uses Gaussian process regression to estimate user demand, it assumes the demand itself is a continuous distribution, whereas the demand in our problem is generated by individual UAVs, and we demonstrate that this demand can be approximated using Gaussian process regression.  
Further, in the coverage problem, the UAVs are assumed to have unlimited capacity and can cover any user demand within a certain range. This is in contrast with our setting, where the MCAs have throughput capacities that must be respected when determining the MCA positions. Thus, our objective is mathematically distinct from~\cite{dist-3D}, and further, the inclusion of the capacity constraints makes the problem even more challenging, necessitating our multi-phase solution.

\section{Conclusion}\label{conclusion}
We have presented a novel distributed solution for collaborative task processing in UAV MEC networks.
One key contribution of this work is to show that the workload can be efficiently approximated by a continuous task field. By adopting this continuum model, we are able to devise a distributed approach to maximize the system throughput; the MCAs adaptively update their positions and the task assignments to maximize the task transmission rates while meeting their processing capacities. Our solution shows up to 28\% improvement over a non-adaptive baseline strategy in experiments. 

This work demonstrates the potential of utilizing continuous models for problems like task offloading, which are traditionally treated as discrete optimization problems. As discussed in Section~\ref{estex.sec}, the accuracy of our task field estimation decreases as the time window $\Delta$ increases. To improve this estimation accuracy, in future work, we plan to investigate more sophisticated estimation techniques that incorporate MSA mobility prediction. In addition, we plan to extend our approach to optimize for the MCA motion energy and use techniques like work stealing to balance workload. Finally, we aim to relax the model to a continuous time approach, rather than a tumbling window approach, so that the MCAs move at independently determined times to optimize for the combination of energy and system throughput.

\section*{Acknowledgment}
This research was partially supported by National Science Foundation Grant CNS-1816307 and  Air Force Office of Scientific Research DDDAS Grant FA9550-19-1-0054.

\begin{appendix}

\section*{Proof of Theorem~\ref{alg1.thm}}
To prove Theorem~\ref{alg1.thm}, we must show that our system model and Alg.~\ref{alg1} satisfy the assumptions needed for convergence of the
projected stochastic gradient descent algorithm as required by Theorem~1 in~\cite{bianchi2012convergence}.
The projected stochastic gradient descent algorithm addresses problems of the form:
\begin{align}
\underset{\theta \in C}{\text{minimize}}~&~\sum_{m=1}^M f_m(\theta).
\end{align}
It is required that $C$ is a non-empty compact convex set and that the functions $f_m$ are (possibly) non-convex and continuously differentiable.
In our setting, $C$ corresponds to $\Omega_c$, which satisfies the requirements. For example, the square region used for $\Omega_c$ in the experiments is non-empty, compact, and convex.
Recall the definition of $f$:
\begin{align}
    f_m(\bu) = \int_{\cV_m} \rho(\bx, t) \lat{\bu_m}{\bx} \der \bx . \label{f2.eq}
\end{align}
We note that $\omega$ is non-decreasing and continuously differentiable.
Thus, as shown in~\cite{du1999centroidal,coverage-ctr}, each $f_m$ is continuously differentiable.

The final requirement to satisfy Theorem~1 in~\cite{bianchi2012convergence} relates to the weight $\xi$ that the MCAs use when averaging their position estimates.
Let $\mathbf{I}$ be the $M \times M$ identity matrix. Further, let~$\mathbf{A}$ be the unweighted adjacency matrix of ${\cal G}_b$, and 
let $\mathbf{D}$ be the~$M \times M$ diagonal matrix where the $(m,m)$th entry is the vertex degree of MCA $m$ in ${\cal G}_b$.
We define the weight matrix~$\mathbf{W} = \mathbf{I} - \xi( \mathbf{D} - \mathbf{A})$, and note that in Alg.~\ref{alg1}, when MCA~$m$ computes a weighted average of the position estimates, it gives weight $\mathbf{W}_{mj}$ to the estimate of MCA~$j$, for each of its neighbors $j$, and it gives weight $\mathbf{W}_{mm}$ to its own estimate.
Theorem~1 in~\cite{bianchi2012convergence} requires that the spectral radius of~$\mathbf{W}$ is strictly less than 1.
We first observe that $\mathbf{W}$ is symmetric and so all of its eigenvalues are real valued.
If $\xi$ is such that $\xi \leq \frac{1}{\delta}$ where $\delta$ is the maximum vertex degree of ${\cal G}_b$, and ${\cal G}_b$ is not bipartite, then it holds that
the eigenvalues of~$\mathbf{W}$ lie strictly between $-1$ and $1$~\cite{anderson85}, and thus, the requirement on the spectral radius of $\mathbf{W}$ is satisfied.
\end{appendix}

\balance
\bibliographystyle{IEEEtran}
\bibliography{reference}

\end{document}